\documentstyle[12pt]{article}
\input{epsf}
\newcommand{\be}{\begin{eqnarray}}
\newcommand{\ee}{\end{eqnarray}}
\newcommand{\ba}{\begin{array}}
\newcommand{\ea}{\end{array}}

\begin{document}
\renewcommand{\thefootnote}{\fnsymbol{footnote}}
%
% BEGIN TITLEPAGE
%

\rightline{\textsl{}} \vspace{0.5cm}
\begin{center}
{\Large Educing GPDs from amplitudes of hard exclusive processes
\footnote{To appear in proceedings of Crimea conference ``NEW TRENDS IN HIGH-ENERGY PHYSICS", Yalta, Sept. 15-22, 2007}}\\
\vspace{0.35cm}
 M.V. Polyakov\\

\vspace{0.35cm}
Petersburg Nuclear Physics
Institute, Gatchina, St.\ Petersburg 188350, Russia\\
and\\
Institut f\"ur Theoretische Physik II,
Ruhr--Universit\"at Bochum, D--44780 Bochum, Germany

%
%\date{\today}
%
\end{center}

\begin{abstract}
The dual parametrization of generalized parton distributions (GPDs) is considered in details.
We discuss which part of information about hadron structure
encoded in GPDs [part of total GPD image]
can be restored from the known amplitude
of a hard exclusive process. The physics content of this partial image is analyzed.

\end{abstract}
\vspace{0.1cm}

\section*{\normalsize \bf Introduction and basics of the dual parametrization of GPDs}
\noindent

The generalized parton distributions (GPDs)
\cite{pioneers}
(for recent reviews of GPDs see Refs.~\cite{GPV,Diehlrev,Belitskyrev}) describe the response of
the target hadron to the well-defined QCD quark and gluon probes of the type:

\be
\label{AtoB}
\langle B| \bar \psi_\alpha(0)\ {\rm P}e^{ig\int_0^z dx_\mu A^\mu}\
\psi_\beta (z) |A\rangle\, ,\qquad
\langle B| G_{\alpha \beta}^a(0)\ \Biggl[{\rm P}e^{ig\int_0^z dx_\mu A^\mu}\Biggr]^{ab}\
G_{\mu \nu}^{b}(z) |A\rangle\, ,
\ee
where the operators are defined on the light-cone, i.e.
$z^2=0$, and $A,B$ are various hadronic states. In this way the hard exclusive processes provide us
with a set of new fundamental probes of the hadronic structure. The hadron image seen by the probes
(\ref{AtoB}) is encoded in the dependence of GPDs on its variables.

 The amplitudes of hard exclusive processes, which we consider as direct observables, are given
by the convolution of GPDs with perturbative kernels. It means that measurements of the amplitudes
provide us with some kind of sectional  images of GPDs -- the convolution integral ``projects out" one
of variables in GPDs.
In this contribution we address the question:
What part of the full hadron image provided by GPDs can be reconstructed
if we know the amplitude of hard exclusive process? This question was recently discussed
in Refs.~\cite{Mueller:2006pm,Kumericki:2007sa,tomography,dieterkrym}. Here we follow the Ref.~\cite{tomography}
adding new ideas and calculations.

For our analysis we employ dual parametrization of GPDs suggested in Ref.~\cite{MaxAndrei}. This parametrization
is based on representation of parton
distributions as an infinite series of $t$-channel exchanges \cite{MVP98}.

In the dual parametrization  the GPD $H(x,\xi,t)$
is expressed in terms of set of functions $Q_{2\nu}(x,t)$ ($\nu=0,1,2,\ldots$).
For the detailed expression see original paper \cite{MaxAndrei}.
We call the functions $Q_{2\nu}(x,t)$ {\it forward-like} because:

\begin{itemize}
\item
At the LO scale dependence of functions $Q_{2\nu}(x,t)$ is given by the standard DGLAP evolution equation, so that these functions
behave as usual parton distributions under QCD evolution.

\item
The function $Q_0(x,t)$ at $t=0$ is related to the forward distribution $q(x)$ as:

\be
Q_0(x,t=0)=\left[ (q(x)+\bar q(x))-\frac x2  \int_x^1 \frac{dz}{z^2}\ ( q(z)+\bar q(z))
\right]\, .
\label{Q0}
\ee
The inverse transformation reads:
\be
q(x)+\bar q(x)= Q_0(x)+\frac{\sqrt x}{2} \int_x^1 \frac{dy}{y^{3/2}}\ Q_0(y)\, .
\ee

\item
The expansion of the GPD $H(x,\xi,t)$ around
the point $\xi=0$ with fixed $x$  to the order $\xi^{2\nu}$ involves
only finite number of functions $Q_{2\mu}(x,t)$ with $\mu\leq \nu$.
For instance, to the order $\xi^2$ these are only $Q_0$ and $Q_2$:

\be
\nonumber
&&H(x,\xi,t)\sim \frac 12\ Q_0(x,t)+\frac{\sqrt x}{4} \int_x^1 \frac{dy}{y^{3/2}}\ Q_0(y,t)\\
\nonumber
&+&\frac{\xi^2}{8}\ \left[-\frac{1-x^2}{x}\frac{\partial}{\partial x} Q_0(x,t)+\frac{1}{8}\int_x^1\frac{dy}{y^3}\ Q_0(y,t)\left(3 \sqrt{\frac{y}{x}}-
\left(\frac{y}{x}\right)^{3/2}\right) \right.\\
\nonumber
&+&\left.
\frac{3}{8}\int_x^1\frac{dy}{y}\ Q_0(y,t)\left(
  \sqrt{\frac{x}{y}}+ \sqrt{\frac{y}{x}}\right) \right. \\
 \label{expansionH}
&+& \left.
\ Q_2(x,t)+\frac{3}{8}\int_x^1\frac{dy}{y}\ Q_2(y,t)\left(
\frac 12\ \sqrt{ \frac{x}{y}}+  \sqrt{\frac{y}{x}}+\frac {5}{2} \left(\frac{y}{x}\right)^{3/2}\right)\right]+O(\xi^4)\,.
\ee
\end{itemize}
The Eq.~(\ref{expansionH}) reveals some of important properties of the forward-like functions.
For example, Eq.~(\ref{expansionH}) dictates that the
forward like functions $Q_{2\nu}(x,t)$ can have small $x$ singularities
of the type $Q_{2\nu}(x,t)\sim \frac{1}{x^{2 \nu+\alpha}}$ where the exponent $\alpha$ governs small $x$ behaviour
of forward quark distribution $q(x)+\bar q(x)\sim 1/x^\alpha$. At first glance such singularities lead to divergency
of of the D-term and
of the Mellin moments of GPDs. However one can easily show that the most singular at
small $x$ terms of $Q_{2\nu}(x,t)$ has the following form:

\be
\label{formofsing}
Q_{2\nu}^{\rm sing}(x,t)=\int_x^1 \frac{dz}{z^{1+2\nu}}\ Q_0(z,t)\ \frac{x^2}{x^2+\epsilon^2}\ R_{\nu}\left(\frac x z\right),
\ee
with
\be
\int_0^1 dz\ z^{k}\ R_{\nu}\left(z\right)=0,~~~~~~~ k\leq 2\nu-1.
\ee
Here we introduce factor containing regularizing parameter $\epsilon$
which is put to zero at the end. Introducing the regularizing factor we modify the parton distribution
in the arbitrarily small vicinity of $x=0$, after performing calculations one has to check that the final result
has a finite limit at $\epsilon\to 0$. The specific form (\ref{formofsing}) of the small-$x$ singular terms of forward like
functions $Q_{2\nu}(x,t)$ (with $\nu\geq 1$) is actually the consequence of the polynomiality of the GPDs Mellin
moments.
The above prescription is very important to understand the naively divergent integrals in our expressions below.
The functions $R_{2\nu}(z)$ can be computed if one knows the expansion of the GPD at small $\xi$:
\be
H(x,\xi,t)=q(x,t)+\xi^2\ f_2(x,t)+\ldots
\ee
For example for the simple model of GPD $H(x,\xi)=q(x)$ we obtain that:
\be
\label{R1}
R_1(z)= z\ \delta^\prime(1-z)-\frac 12\ \delta(1-z)-\left(1+3 z \right)\, .
\ee
We shall see below that the amplitudes of the hard exclusive processes are expressed in terms of
single function
\be
N(x,t)=\sum_{\nu=0}^{\infty} x^{2\nu} Q_{2\nu}(x,t)\, .
\label{Nifunction}
\ee
Its small $x$ behaviour has  the following form

\be \label{Nsmallx} N(x,t)\sim \frac{N_0}{x^\alpha}+\ldots \, ,
\ee with $\alpha<2$. Important to note that neither the exponent
$\alpha$ nor the coefficient $N_0$ in front of $1/x^\alpha$ is
determined by the forward parton distributions. However, as we
discussed above, the singular terms, which deviate from those
dictated by the forward parton distributions  \footnote{If the forward
distribution behaves $q(x)\sim 1/x^\alpha$ the corresponding peace
has the following behaviour $N(x,t)\sim \frac{\alpha+\frac
12}{\alpha+1}\ 1/x^\alpha$} have very specific form (\ref{Nsing}).
In the first phenomenological applications of the dual
parametrization to describe the DVCS data \cite{Guzey} this was
not taken into account (it was assumed that $Q_{2\nu}(x)\sim
1/x^\alpha$). The inclusion of the singular terms of the type
(\ref{formofsing}) into the parametrization can considerably
change the results and conclusions of these papers.

\section*{\normalsize \bf Forward-like distributions in terms of amplitudes}
\noindent

The leading order amplitude of hard exclusive reactions is expressed
in terms of the following elementary amplitude\footnote{We restrict ourselves to the
singlet (even signature) amplitudes, generalization for odd signature amplitudes is trivial.}:
\be
A(\xi,t)=\int_{0}^1 dx\,H(x,\xi,t)\,
\left[\frac 1{\xi-x-i0} -\frac 1{\xi+x-i0}\right].
\label{elementaryAMP}
\ee
We see that the amplitude is given by the convolution integral in which dependence
of GPDs on variable $x$ is ``integrated out". Mathematically from the equations
(\ref{elementaryAMP}) one can not completely restore
the GPD $H(x,\xi,t)$. So we are not able to perform ``complete imaging" of the target hadron
from the knowledge of the amplitude.
The key question is: what part of the ``complete image" can be restored from the known amplitude?
What is the physics content of the restorable part of the complete image?
Attempt to answer these questions in the framework of the dual parametrization was done in Ref.~\cite{tomography}.
We refer the reader to this paper for details, here we just give main results and more detailed discussion
of subtleties.

We can express the amplitudes in terms of forward-like function (see Eq.~(\ref{Nifunction}))
$N(x,t)$ as following \cite{MaxAndrei}:
\begin{eqnarray}
\nonumber
{\rm Im\ } A(\xi,t)&=&
\int_{\frac{1-\sqrt{1-\xi^2}}{\xi}}^1
\frac{dx}{x} N(x,t)\
\Biggl[
\frac{1}{\sqrt{\frac{2 x}{\xi}-x^2-1}}
\Biggr]\, ,\\
\nonumber
{\rm Re\ } A(\xi,t)&=&
\int_0^{\frac{1-\sqrt{1-\xi^2}}{\xi}}
\frac{dx}{x} N(x,t)\
\Biggl[
\frac{1}{\sqrt{1-\frac{2 x}{\xi}+x^2}} +
\frac{1}{\sqrt{1+\frac{2 x}{\xi}+x^2}}-\frac{2}{\sqrt{1+x^2}}
\Biggr]  \\
&+&\int^1_{\frac{1-\sqrt{1-\xi^2}}{\xi}}
\frac{dx}{x} N(x,t)\
\Biggl[
\frac{1}{\sqrt{1+\frac{2 x}{\xi}+x^2}}-\frac{2}{\sqrt{1+x^2}}
\Biggr]+ 2 D(t)
\, .
\label{REIM}
\end{eqnarray}
Here we introduced the D-form factor:
\be
D(t)=\sum_{n=1}^\infty d_n(t)=\frac 12 \int_{-1}^1 dz\ \frac{D(z,t)}{1-z}\, ,
\ee
where $D(z,t)$ is the D-term \cite{PW99}\footnote{The coefficients  $d_n(t)$ individually can be obtained
from the generating function \cite{tomography}:
\be
\label{generatingd}
\sum_{n=1\atop \scriptstyle{\rm odd}}^{\infty}\ d_n(t)\ \alpha^n =
\frac{1}{\alpha}\ \int_0^1\ \frac{dz}{z}\ \sum_{\nu=0}^\infty
(\alpha\ z)^{2\nu} Q_{2\nu}(z,t) \left(\frac{1}{\sqrt{1+\alpha^2 z^2}}-\delta_{\nu 0}
\right)
\ee
}.

Now we clearly see that the knowledge of  the LO amplitude is equivalent to
the knowledge of the function $N(x,t)$ and D-form factor $D(t)$. Moreover the D-form factor
can be computed in terms of $N(x,t)$ and $Q_0(x,t)$. Note that the latter function
is to great extend is fixed by the forward parton distributions, see Eq.~(\ref{Q0}).
The expression for the D-form factor is the following \cite{tomography}:

\be
D(t) =
 \int_0^1\ \frac{dz}{z}\
Q_{0}(z,t) \left(\frac{1}{\sqrt{1+ z^2}}-1
\right)+
 \int_0^1\ \frac{dz}{z}\
\left[N(z,t)-Q_{0}(z,t)\right] \ \frac{1}{\sqrt{1+ z^2}}
\label{DFFNQ0}
\ee
Given the small $x$ behaviour (\ref{Nsmallx}) of the function $N(x,t)$ naively one may think
that the above integral is divergent. However it is not the case due to the very specific form
of the singular terms, see Eq.~(\ref{formofsing}). Because of this specific form of the
small $x$ singularities we can represent the function
$N(x,t)$ in the following way:

$$
N(x,t)=Q_0(x,t)+ \widetilde N^{\rm reg}(x,t)+\widetilde N^{\rm sing}(x,t)\, ,
$$
where the function $\widetilde N^{\rm reg}(x,t)$ goes to zero as $x\to 0$ whereas
the singular part behaves as $\sim 1/x^\alpha$ and has the following structure:

\be
\label{Nsing}
\widetilde N^{\rm sing}(x,t)=\int_x^1 \frac{dz}{z}\ Q_0(z,t)\ \frac{x^2}{x^2+\epsilon^2}\ R\left(\frac x z\right)\, ,
\ee
where the function $R(z)$ has the property:

\be
\label{Rproperty}
\int_0^1 dz \frac{1}{z} \ R(z)=0.
\ee
Now we can rewrite the representation for the D-form factor (\ref{DFFNQ0}) as follows:

\be
D(t) =
 \int_0^1\ \frac{dz}{z}\
\left[Q_{0}(z,t)+\widetilde N^{\rm sing}(z,t)\right] \left(\frac{1}{\sqrt{1+ z^2}}-1
\right)+
 \int_0^1\ \frac{dz}{z}\
\widetilde N^{\rm reg}(x,t)\ \frac{1}{\sqrt{1+ z^2}}.
\label{DFFNQ0reg}
\ee
Here all integrals are explicitly convergent and therefore one can safely put $\epsilon=0$.
It implies that, for practical calculations, one can use the following prescription. If one knows the small $x$
behaviour of the function $N(x,t)-Q_0(x,t)$\footnote{We see below that it is equivalent to the knowledge of the small
$\xi$ behaviour of the amplitude which can be directly measured and small $x$ asymptotic of the forward parton distribution}:

\be
\label{asymp}
N(x,t)-Q_0(x)\sim \sum_i \frac{A_i}{x^\alpha_i}+\ldots,\ \ \ \ {\rm with}\ 0<\alpha_i<2
\ee
then we can define the singular part of the function $N(x,t)$ in the following way:

\be
\label{Nsing2}
\widetilde N^{\rm sing}(x,t)=\sum_i A_i \frac{1}{x^{\alpha_i}} \
\frac{\int_x^1 dz\ z^{\alpha_i-1} R(z )}{\int_0^1 dz z^{\alpha_i-1} R(z)}\, ,
\ee
with {\it arbitrary } function $R(z)$ which satisfies Eq.~(\ref{Rproperty}). Now it is obvious that
the function defined as:
\be
\widetilde N^{\rm reg}(x,t)=N(x,t)-Q_0(x,t)-\widetilde N^{\rm sing}(x,t)\, ,
\ee
tends to zero if $x\to 0$.

As it is shown in Ref.~\cite{tomography} the Mellin moments of $N(x,t)$
allow to fix the angular momentum of exchanged partons.

\be
\int_0^1 dx\ x^{J-1}\ N(x,t)=\frac 12 \int_{-1}^1 dz \frac{\Phi_J(z,t)}{1-z}\, ,
\label{Jexchange}
\ee
where $\Phi_J(z,t)$ is the distribution amplitude corresponding to two quark exchange in the $t$-channel
with fixed angular momentum $J$. The quantity on RHS of Eq.~(\ref{Jexchange}) carries valuable information
about the hadron structure -- it tells how the target nucleon responses to the
well defined string-like quark-antiquark probe (see Eq.~(\ref{AtoB})) with fixed angular momentum $J$.

We spent some time to discuss how to compute the D-form factor if one knows the functions $N(x,t)$
and $Q_{0}(x,t)$. The latter function is closely related to the forward parton distributions
and the former can be computed directly from the observed amplitudes. The final result for
the inversion problem is the following (see details \cite{tomography}):

\be
N(x,t)=\frac{2}{\pi}\ \frac{x(1-x^2)}{~~~(1+x^2)^{3/2}} \int_\frac{2 x}{1+x^2}^1\frac{d\xi}{\xi^{3/2}}\
\frac{1}{\sqrt{\xi-\frac{2 x}{1+x^2}}} \left\{ \frac 12{\rm Im\ } A(\xi,t)-\xi \frac{d}{d\xi}{\rm Im\ } A(\xi,t)
\right\}
\label{nina}
\ee
This remarkable formula allows to restore the function $N(x,t)$ from the measured imaginary part
of the amplitude. Note that the inversion formula contains the amplitude only in the physical region.
The Eq.~(\ref{nina}) directly translates any knowledge (theoretical, model or experimental) of the amplitude
into comprehension of GPDs.
At $\xi\to 1$ the imaginary part of the amplitude should go to zero. Let us assume that
$
{\rm Im\ } A(\xi,t) \sim (1-\xi)^\beta
$ as $\xi\to 1$, than from Eq.~(\ref{nina}) one can easily obtain that

$$
N(x,t)\sim \frac{1}{2^{\beta-1}} \frac{\Gamma\left(\beta+1\right)\Gamma\left(\frac 12\right)}{\Gamma\left(\beta+\frac 12\right)}\ (1-x)^{2 \beta}
$$
as $x\to 1$.

As to small $x$ behaviour, the Regge like asymptotic of the imaginary part of the amplitude
${\rm Im\ } A(\xi,t) \sim 1/\xi^\alpha$, according to Eq.~(\ref{nina}), corresponds to the following
small $x$ behaviour of $N(x,t)$:

$$
N(x,t)\sim \frac{1}{2^\alpha}\ \frac{\Gamma(1+\alpha)}{\Gamma(\frac 12)\Gamma(\frac 12+\alpha)}\ \frac{1}{x^\alpha}\, .
$$

Up to now we made use of only the imaginary part of the amplitude, let us study the real part of the amplitude.
For this we take the expression (\ref{nina}) for the $N(x,t)$ in terms of ${\rm Im} A(\xi,t)$ and substitute
it into Eq.~(\ref{REIM}). After some simple calculations we arrive to the following expression for real part
of the amplitude:

\be
{\rm Re} A(\xi,t)= 2 D(t)+\frac{1}{\pi}\ vp \int_0^1 d\zeta \ {\rm Im} A(\zeta,t) \left(\frac{1}{\xi-\zeta}-
\frac{1}{\xi+\zeta} \right)\, ,
\label{DR}
\ee
in which we can immediately recognize the dispersion relation for the amplitude with one subtraction
at non-physical point $\xi=\infty$ (corresponding to $\nu=(s-u)/4m=0$). The D-form factor is the corresponding
subtraction constant. This result was obtained recently in Refs.~\cite{Teryaev,DiehlIvanov} by independent methods.
We see that the dual parametrization automatically ensures the dispersion relations for the amplitudes.

The very idea of the dual parametrization of GPDs in terms of $t$-channel exchanges was motivated by the crossing
relations
\cite{MVP98} between GPDs and generalized distribution amplitude \cite{DiehlTeryaev}. The later enter the description
of the hard exclusive processes in the cross channel, like $\gamma^*+\gamma\to h+\bar h$. In the LO
the amplitude of the cross process can be expressed in terms of the function $N(x,t)$ analytically continued
to time-like $t$ ($t>0$):

\be
A^{\rm cross}(\eta,t)=
\int_0^{1}
\frac{dx}{x} N(x,t)\
\Biggl[
\frac{1}{\sqrt{1-2 x \eta+x^2}} +
\frac{1}{\sqrt{1+ 2 x\eta+x^2}}-\frac{2}{\sqrt{1+x^2}}
\Biggr]  + 2 D(t)\, .
\ee
Here $-\eta$ is directly related  to the $\cos(\theta_{cm})$-- cosine of scattering angle in centre of mass system,
see for details Refs.~\cite{DiehlTeryaev,MVP98}. Now substituting our inversion formula (\ref{nina}) into this expression
we obtain, rather simple result:

\be
A^{\rm cross}(\eta,t)=\frac 2\pi \int_0^{|\eta|} d\xi \frac{\xi}{1-\xi^2}\ {\rm Im} A\left(\frac {\xi}{|\eta|},t\right)
+2\ D(t)\, .
\ee
Actually this equation is the consequence of the dispersion relation (\ref{DR}).

Now we have all ingredients to make proposals for possible ways to analyze date on hard exclusive reactions.

\section*{\normalsize \bf Possible ways to comprehend GPDs from data}
\noindent

The observables of hard exclusive processes can be analyzed assuming certain functional form of ${\rm Im} A(\xi,t)$ with set of free parameters.
It can be the form with motivated by some
theories (like Regge-motivated \cite{jenk}, etc.) or one can choose some flexible enough functions. For instance one can take as the functional form:

\be
\label{fitf}
{\rm Im} A(\xi,t)={\rm Im} A_0(\xi,t)+\sum_i\ \frac{P_i}{x^{\alpha_i}}\ (1-x)^{\beta_i}\, ,
\ee
where $P_i,\alpha_i, \beta_i$ is the set of free parameters which should be adjusted to data, ${\rm Im} A_0(\xi,t)$ is the amplitude computed
with the first Eq.~(\ref{REIM}) taking $N(x)=Q_0(x)$, i.e. it is fixed by the forward parton distribution.
The reader can select any other function form of the amplitude. With help of our master formula (\ref{nina}) we can immediately compute the function
$N(x,t)$, for instance for the choice (\ref{fitf}):

\be
N(x,t)=Q_0(x,t)+ \sum_i P_i\ F^{[\alpha_i,\beta_i]}(x,t),
\ee
with $F^{[\alpha_i,\beta_i]}(x,t)$ given by the simple combination of hypergeometric functions. Knowing the function $N(x,t)$ and the forward parton distribution
$Q_0(x,t)$ we can calculate the real part of the amplitude and the $D$ form factor. To compute the $D$ form factor we should figure out the small $x$ asymptotic
(\ref{asymp}) and construct $\widetilde N^{\rm sing}(x,t)$ according to Eq.~(\ref{Nsing2}). For the amplitude of the form (\ref{fitf}) we obtain:

\be
\widetilde N^{\rm sing}(x,t)&=& \sum_i P_i \frac{\Gamma(1+\alpha_i)}{2^{\alpha_i} \sqrt{\pi} \Gamma(\alpha_i+\frac 12)}\cdot
\left\{\frac{1}{x^{\alpha_i}} \frac{\int_x^1 dz\ z^{\alpha_i-1} R(z )}{\int_0^1 dz z^{\alpha_i-1} R(z)}-\right.
\nonumber\\
&-&
\left.
\frac{\beta_i (2\alpha_i-1)}{\alpha_1}
\frac{1}{x^{\alpha_i-1}} \frac{\int_x^1 dz\ z^{\alpha_i-2} R(z )}{\int_0^1 dz z^{\alpha_i-2} R(z)}\
\right\}\, ,
\ee
with arbitrary function $R(z)$ possessing the property (\ref{Rproperty}). After one can compute the $D$ form factor (\ref{DFFNQ0reg}) and the real part
of the amplitude (\ref{REIM}) in terms of the adjustable parameters which enter the chosen functional form for the imaginary part of the amplitude.
In principle, these adjustable parameters can be determined from the data on imaginary part of the amplitude (like beam spin asymmetry for DVCS), then
the real part and the D- term is predicted. Again let us give the result for our choice of the fitting function (\ref{fitf}). For such choice the
$D$ form factor is:

\be
D(t)=\int_0^1\ \frac{dz}{z}\
Q_{0}(z,t) \left(\frac{1}{\sqrt{1+ z^2}}-1
\right)+ \frac{1}{\pi} \sum_i\ P_i\ \frac{\Gamma(-\alpha_i)\ \Gamma(1+\beta_i)}{\Gamma(1+\beta_i-\alpha_i)}\, .
\ee

Till now we assumed that the forward limit of the GPD at hand is known. However in some cases
the forward limit is not known and should be determined. The most interesting example is the
nucleon GPD $H_{M}(x,\xi,t)=H(x,\xi,t)+E(x,\xi,t)$\footnote{This ``magnetic-like"
GPD has correct angular momentum quantum numbers in the $t$-channel.
The full classification of such nucleon GPD is given in \cite{Diehlrev}.}.
Its forward limit is given by unknown function $Q_0^{(M)}(x,t)$ which is normalized to the angular momentum carried by quark in the nucleon $J^Q$:

\be
\int_0^1 dx\ x\ Q_0^{(M)}(x,t)=\frac 5 3\ J^Q(t)\, .
\ee
Actually it is one of the main goals of measurements of hard exclusive processes to determine this function.

Well, as we do not know the forward function $Q_0^{(M)}(x,t)$ there is no
point to single out its contribution to the imaginary part of the amplitude.
Instead, we simply parametrize the
${\rm Im}\ A^{(M)}$\footnote{We keep the superscript $M$ to stress
that we consider the case with unknown forward limit} as the whole, e.g.
by the following function:

\be
\label{fitf2}
{\rm Im} A^{(M)}(\xi,t)=\sum_i\ \frac{P_i}{x^{\alpha_i}}\ (1-x)^{\beta_i}\,.
\ee
Now again we can easily perform analysis along the line discussed above,
the only difference is that the expression for the
D-form factor is different:

\be
\label{dterm2}
D(t)=-\int_0^1\ \frac{dz}{z}\
\left[ Q_{0}^{(M)}(z,t)-Q_{0}^{(M){\rm sing}}(z,t)\right]+ \frac{1}{\pi} \sum_i\ P_i\ \frac{\Gamma(-\alpha_i)\ \Gamma(1+\beta_i)}{\Gamma(1+\beta_i-\alpha_i)}\, .
\ee
Here  $Q_{0}^{(M)}(x,t)$ is unknown (to be determined) forward parton distribution and $Q_{0}^{(M) {\rm sing}}(x,t)$ its singular part computed
according to the Eq.~(\ref{Nsing2}) with obvious replacement of functions.
Now we can parametrize the unknown function $Q_{0}^{(M)}(x,t)$ and vary its parameters to achieve
the agreement with data. In the case of nucleon GPD $H_{M}(x,\xi,t)=H(x,\xi,t)+E(x,\xi,t)$ we know
that corresponding $D$-form factor is zero and we can constrain
the forward function $Q_{0}^{(M)}(x,t)$ from the following equation:

\be
\int_0^1\ \frac{dz}{z}\
\left[ Q_{0}^{(M)}(z,t)-Q_{0}^{(M){\rm sing}}(z,t)\right]= \frac{1}{\pi} \sum_i\ P_i\ \frac{\Gamma(-\alpha_i)\ \Gamma(1+\beta_i)}{\Gamma(1+\beta_i-\alpha_i)}\, .
\ee
Here the parameters $P_i,\alpha_i$ and $\beta_i$ are determined from fitting the
imaginary part of the amplitude\footnote{The reader can easily derive corresponding equation
for her/his beloved functional form of the amplitude.
As a curiosity :) we note that if we assume that forward limit of ``magnetic" nucleon GPD has the form:

$$
Q_{0}^{(M)}(x,t)=\frac{C}{x^a}(1-x)^b\, ,
$$
we can obtain $J^Q$ from the fit to data on ${\rm Im} A^{(M)}(\xi,t)$ by the functional form (\ref{fitf2}) as follows:

\be J^Q =\frac{3 a (a-1)}{5\pi (2+b-a)(1+b-a)}\ \sum_i\ P_i\
\frac{\Gamma(-\alpha_i)\
\Gamma(1+\beta_i)}{\Gamma(1+\beta_i-\alpha_i)}\, . \ee }. Note
that the expression in Eq.~(\ref{dterm2}) corresponds to the
$D$-form factor computed from the amplitude (\ref{fitf2}) with
help of analytical regularization discussed in
Refs.~\cite{Mueller:2006pm,Kumericki:2007sa,dieterkrym}.

Up to now we discussed possibilities to fit data at fixed photon virtuality $Q^2$.
If we are interested in studies of the scaling violation
and determination of the whole GPD we have to model the forward-like functions $Q_{2\nu}(x,t)$ individually. In this case one
has more freedom and one needs more a priori input.

For modelling of forward-like function we propose the following general ansatz:

\be
\label{ansatz}
Q_{2\nu}(x,t)=\int_x^1\ \frac{dz}{z}\ Q_0(z,t)\
P_{\nu}\left(\frac x z\right)+\frac{1}{x^{2\nu}}\int_x^1\ \frac{dz}{z}\
Q_0\left(\frac{x}{z},t\right)\ z^{2\nu} R_{\nu}\left( z\right)\, .
\ee
Here the second terms accounts for most singular small $x$ asymptotic of the forward-like functions. As we discussed
above, the specific structure of
these singularities (\ref{formofsing}) ensures that $R_\nu(z)$ satisfies the condition:
\be
\label{condition}
\int_0^1 \frac{dz}{ z^{2\nu-k}}\ z^{2\nu} R_{\nu}\left(z\right)=0,~~~~~~~ k=1,3,\ldots,2\nu-1.
\ee
Possible ways to model the functions $P_{\nu}(z)$ in Eq.~(\ref{ansatz}) were discussed in Ref.~\cite{tomography}.
As to modelling functions $R_\nu(z)$ one may try the following form:

\be
R_\nu(z)=w(z)\sum_{j=0}^{2\nu-1}\ r_j\ p_{j}(z),
\ee
where $p_j(z)$ is set of orthogonal on the interval $[0,1]$ with the weight $w(z)$ polynomials. This form
automatically satisfies the condition (\ref{condition}). Alternatively one can choose to model the functions
$R_\nu(z)$ allowing the presence of $\delta$-functions, like in Eq.~(\ref{R1}).

We see that there there is rather large freedom in choice of the functional form for the forward-like
functions $Q_{2\nu}(x)$. Further constraints for possible form and size of these functions we can obtain
computing the forward-like functions in various models for GPDs, like, for instance, chiral quark-soliton model
\cite{chqsm}.

\section*{\normalsize \bf Further possible uses of dual parametrization}
\noindent

Here, instead of conclusion, we outline couple of ideas about use of dual parametrization apart
from those which obviously follow from the discussion above.

One of advantages of the dual parametrization is that it allows to separate the contribution
of the usual forward parton distributions to the amplitude of a hard exclusive process. This is
the function $Q_0(x,t)$ in the function $N(x,t)=Q_0(x,t)+x^2\ Q_2(x,t)+\ldots $, the latter completely
determines the amplitude. This feature allows us to study ``truly non-forward" effects in the hard
processes. For example, studying  DVCS on nuclei, with help of dual parametrization, we can easily
single out ``non-forward EMC effect". The usual EMC effect is taken into account by the
function $Q_0(x,t)$ which is determined by nuclear forward parton distributions. Model calculations
\cite{sily,marat} predict unusual behaviour of $x$-moment of the function $Q_2(x,t)$ with
atomic number. The dual parametrization allows to reveal such new nuclear effects from the data.

Another possible application of the dual parametrization is related to Eq.~(\ref{Jexchange}). Apart
from obvious uses of the Eq.~(\ref{Jexchange}), it can bring new insight into spectroscopy of baryon resonances.
One can repeat the discussion above for the ``non-diagonal" GPDs which are given by the matrix elements
of the light-cone operators (\ref{AtoB}) for transitions between nucleon and meson-nucleon states\footnote{Such GPDs enter
the description of hard exclusive processes like $\gamma^*+N\to \gamma+ ({\rm meson}\ N)$. }.
in this case one can introduce the function $N(x,t,\ldots)$ which in this case depends on additional
variables (denoted by $\ldots$, the definition of these variables and their relation to partial wave decomposition
of the final meson-nucleon system can be found in Ref.~\cite{pingpd}) which characterize the final meson-nucleon state
(its invariant mass, orbital momentum, etc.). The corresponding function can be obtained from data
using the Eq.~(\ref{nina}). Now if one computes the $x^{J-1}$ moment of this function, one obtains,
according to Eq.~(\ref{Jexchange}), the meson-nucleon state produced by the well defined probe with spin $J$.
This allows to excite new baryon resonances which couple weakly to standard probes like photons, also it allows
to study the properties of known resonances comparing their excitations by probes with different spins.

\section*{\normalsize\bf Acknowledgements}
We are thankful to  N.~Kivel, D.~M\"uller and K.~Semenov-Tian-Shansky for many
valuable discussions.
The
work is supported
by the Sofja Kovalevskaja Programme of the Alexander von Humboldt
Foundation, the Federal Ministry of Education and Research and the
Programme for Investment in the Future of German Government.


\begin{thebibliography}{99}
\bibitem{pioneers}
D. M\"uller, D. Robaschik, B. Geyer, F.M. Dittes, and J. Horejsi,
Fortschr.~Phys. {\bf 42}, 101 (1994);\\
%%CITATION = HEP-PH 9812448;%%
X.~D.~Ji,
  %``Gauge invariant decomposition of nucleon spin,''
  Phys.\ Rev.\ Lett.\  {\bf 78} (1997) 610
  [arXiv:hep-ph/9603249];\\
  %%CITATION = PRLTA,78,610;%%
   A.~V.~Radyushkin,
  %``Scaling Limit of Deeply Virtual Compton Scattering,''
  Phys.\ Lett.\  B {\bf 380} (1996) 417
  [arXiv:hep-ph/9604317];\\
  %%CITATION = PHLTA,B380,417;%%
X.~D.~Ji,
  %``Deeply-virtual Compton scattering,''
  Phys.\ Rev.\  D {\bf 55} (1997) 7114
  [arXiv:hep-ph/9609381].
  %%CITATION = PHRVA,D55,7114;%%

\bibitem{GPV}
K.~Goeke, M.~V.~Polyakov and M.~Vanderhaeghen,
%``Hard exclusive reactions and the structure of hadrons,''
Prog.\ Part.\ Nucl.\ Phys.\  {\bf 47} (2001) 401
[arXiv:hep-ph/0106012].
%%CITATION = HEP-PH 0106012;%%
\bibitem{Diehlrev}
  M.~Diehl,
  %``Generalized parton distributions,''
  Phys.\ Rept.\  {\bf 388} (2003) 41
  [arXiv:hep-ph/0307382].
  %%CITATION = PRPLC,388,41;%%

\bibitem{Belitskyrev}
  A.~V.~Belitsky and A.~V.~Radyushkin,
  %``Unraveling hadron structure with generalized parton distributions,''
  Phys.\ Rept.\  {\bf 418} (2005) 1
  [arXiv:hep-ph/0504030].
  %%CITATION = PRPLC,418,1;%%




\bibitem{Mueller:2006pm}
  D.~Mueller,
  {\it ``Pomeron dominance in deeply virtual Compton scattering and the femto
  holographic image of the proton,''}
  arXiv:hep-ph/0605013.
  %%CITATION = HEP-PH/0605013;%%
\bibitem{Kumericki:2007sa}
  K.~Kumericki, D.~Muller and K.~Passek-Kumericki,
  {\it``Towards a fitting procedure for deeply virtual Compton scattering at
  next-to-leading order and beyond,''}
  arXiv:hep-ph/0703179.
  %%CITATION = HEP-PH/0703179;%%

  \bibitem{tomography}
  M.~V.~Polyakov,
  {\it ``Tomography for amplitudes of hard exclusive processes,''}
  arXiv:0707.2509 [hep-ph].
  %%CITATION = ARXIV:0707.2509;%%
\bibitem{dieterkrym}
  K.~Kumericki, D.~Muller and K.~Passek-Kumericki,
  {\it ``Fitting DVCS at NLO and beyond,''}
  arXiv:0710.5649 [hep-ph].
  %%CITATION = ARXIV:0710.5649;%%


\bibitem{MaxAndrei}
  M.~V.~Polyakov and A.~G.~Shuvaev,
  {\it ``On 'dual' parametrizations of generalized parton distributions,''}
  arXiv:hep-ph/0207153.
  %%CITATION = HEP-PH/0207153;%%

\bibitem{MVP98}
M.V.~Polyakov,
Nucl.~Phys. {\bf B555}, 231 (1999).
%%CITATION = HEP-PH 9809483;%%
\bibitem{Guzey}
  V.~Guzey and M.~V.~Polyakov,
  %``Dual parameterization of generalized parton distributions and  description
  %of DVCS data,''
  Eur.\ Phys.\ J.\  C {\bf 46}, 151 (2006)
  [arXiv:hep-ph/0507183];\\
  %%CITATION = EPHJA,C46,151;%%
 V.~Guzey and T.~Teckentrup,
  %``The dual parameterization of the proton generalized parton distribution
  %functions H and E and description of the DVCS cross sections and
  %asymmetries,''
  Phys.\ Rev.\  D {\bf 74}, 054027 (2006)
  [arXiv:hep-ph/0607099].
  %%CITATION = PHRVA,D74,054027;%%


\bibitem{PW99}
M.V. Polyakov and C. Weiss, Phys.~Rev.~D {\bf 60}, 114017 (1999).
%%CITATION = HEP-PH 9902451;%%


\bibitem{sily}
  M.~V.~Polyakov,
  %``Generalized parton distributions and strong forces inside nucleons and
  %nuclei,''
  Phys.\ Lett.\  B {\bf 555}, 57 (2003)
  [arXiv:hep-ph/0210165].
  %%CITATION = PHLTA,B555,57;%%


\bibitem{Teryaev}
 O.~V.~Teryaev,
  {\it ``Analytic properties of hard exclusive amplitudes,''}
  arXiv:hep-ph/0510031;\\
  %%CITATION = HEP-PH/0510031;%%
  I.~V.~Anikin and O.~V.~Teryaev,
  %``Dispersion relations and subtractions in hard exclusive processes,''
  arXiv:0704.2185 [hep-ph].
  %%CITATION = ARXIV:0704.2185;%%

\bibitem{DiehlIvanov}
  M.~Diehl and D.~Y.~Ivanov,
  %``Dispersion representations for hard exclusive processes,''
  arXiv:0707.0351 [hep-ph].
  %%CITATION = ARXIV:0707.0351;%%


\bibitem{DiehlTeryaev}
M.~Diehl, T.~Gousset, B.~Pire and O.~Teryaev,
%``Probing partonic structure in gamma* gamma $\to$ pi pi near threshold,''
Phys.\ Rev.\ Lett.\  {\bf 81} (1998) 1782
[arXiv:hep-ph/9805380].
%%CITATION = HEP-PH 9805380;%%


\bibitem{jenk}
  L.~L.~Jenkovszky, V.~K.~Magas and A.~Vall,
  %``Generalized parton distributions: A pragmatic approach,''
  Phys.\ Part.\ Nucl.\  {\bf 36} (2005) S152.
  %%CITATION = PPNUE,36,S152;%%

\bibitem{marat}
  V.~Guzey and M.~Siddikov,
  %``On the A-dependence of nuclear generalized parton distributions,''
  J.\ Phys.\ G {\bf 32}, 251 (2006)
  [arXiv:hep-ph/0509158].
  %%CITATION = JPHGB,G32,251;%%


\bibitem{chqsm}
  V.~Y.~Petrov, P.~V.~Pobylitsa, M.~V.~Polyakov, I.~Bornig, K.~Goeke and C.~Weiss,
  %``Off-forward quark distributions of the nucleon in the large N(c) limit,''
  Phys.\ Rev.\  D {\bf 57} (1998) 4325
  [arXiv:hep-ph/9710270];\\
  %%CITATION = PHRVA,D57,4325;%%
 M.~Penttinen, M.~V.~Polyakov and K.~Goeke,
  %``Helicity skewed quark distributions of the nucleon and chiral symmetry,''
  Phys.\ Rev.\  D {\bf 62} (2000) 014024
  [arXiv:hep-ph/9909489];\\
  %%CITATION = PHRVA,D62,014024;%%
  J.~Ossmann, M.~V.~Polyakov, P.~Schweitzer, D.~Urbano and K.~Goeke,
  %``The generalized parton distribution function (E(u)+E(d))(x,xi,t) of the
  %nucleon in the chiral quark soliton model,''
  Phys.\ Rev.\  D {\bf 71} (2005) 034011
  [arXiv:hep-ph/0411172].
  %%CITATION = PHRVA,D71,034011;%%

\bibitem{pingpd}
M.~V.~Polyakov,
  {\it ``$N \to \Delta$ and $N \to N \pi$ DVCS and skewed quark distributions,''}
%\href{http://www.slac.stanford.edu/spires/find/hep/www?irn=4502647}{SPIRES entry}
Prepared for 8th International Conference on the Structure of Baryons (Baryons 98), Bonn, Germany, 22-26 Sep 1998;\\
  M.~V.~Polyakov and S.~Stratmann,
  {\it ``Soft pion emission in hard exclusive pion production,''}
  arXiv:hep-ph/0609045.
  %%CITATION = HEP-PH/0609045;%%

\end{thebibliography}
\end{document}